\documentclass{elsart}
\usepackage{graphicx,amssymb}

\begin{document}

\begin{frontmatter}


\title{Phase transitions of the mixed spin-1/2 and spin-$S$ Ising model 
on a three-dimensional decorated lattice with a layered structure\thanksref{grant}}       
\thanks[grant]{The financial support provided under the grant Nos.~VEGA~1/0128/08 
and VVGS~23/07-08 is gratefully acknowledged. This work was partially supported 
also by the Slovak Research and Development Agency under the contract LPP-0107-06.}
\author[UPJS]{Jozef Stre\v{c}ka}, 
\ead{jozef.strecka@upjs.sk}
\ead[url]{http://158.197.33.91/$\thicksim$strecka}
\author[UPJS]{J\'an Dely}, and
\author[TUKE]{Lucia \v{C}anov\'a}
\address[UPJS]{Department of Theoretical Physics and Astrophysics, 
Faculty of Science, \\ P. J. \v{S}af\'{a}rik University, Park Angelinum 9,
040 01 Ko\v{s}ice, Slovak Republic}
\address[TUKE]{Department of Applied Mathematics, Faculty of Mechanical Engineering, \\ 
Technical University, Letn\'a 9, 042 00 Ko\v{s}ice, Slovak Republic}
\journal{Physica A}
\begin{abstract}
Phase transitions of the mixed spin-1/2 and spin-$S$ ($S \geq 1/2$) Ising model on a three-dimensional (3D) decorated lattice with a layered magnetic structure are investigated within the framework 
of a precise mapping relationship to the simple spin-1/2 Ising model on the tetragonal lattice. 
This mapping correspondence yields for the layered Ising model of mixed spins plausible results 
either by adopting the conjectured solution for the spin-1/2 Ising model on the orthorhombic 
lattice [Z.-D.~Zhang, Philos.~Mag.~87~(2007)~5309-5419] or by performing extensive Monte Carlo 
simulations for the corresponding spin-1/2 Ising model on the tetragonal lattice. 
It is shown that the critical behaviour markedly depends on a relative strength of axial 
zero-field splitting parameter, inter- and intra-layer interactions. The striking spontaneous 
order captured to the 'quasi-1D' spin system is found in a restricted region of interaction 
parameters, where the zero-field splitting parameter forces all integer-valued decorating 
spins towards their 'non-magnetic' spin state. 
\end{abstract}
\end{frontmatter}

\section{Introduction}
\label{intro}

Phase transitions and critical phenomena of rigorously solvable interacting many-particle 
systems are much sought after in the modern equilibrium statistical mechanics as 
they offer valuable insight into a cooperative nature of phase changes \cite{baxt82}.  
Beside this, the usefulness of mathematically tractable models can also be viewed
in providing guidance on a reliability of various approximative techniques, which 
are often needed for treating more complicated models that preclude exact analytical 
treatment. \textit{Decorated planar Ising models}, which can be constructed by adding 
one or more spins on bonds of some original lattice, belong to the simplest mathematically 
tractable lattice-statistical models (see Ref.~\cite{syoz72} and references cited therein). 
The main advantage of decorated Ising models consists in a relative simple way 
of obtaining their exact solutions. As a matter of fact, several decorated planar Ising 
models can straightforwardly be solved by employing the generalized decoration-iteration 
transformation \cite{fish59,roja09} that relates their exact solution to that one of 
the simple spin-1/2 Ising model on a corresponding undecorated lattice, which is generally 
known for many planar lattices of different topologies \cite{utiy51,domb60,lin86}.

Quite recently, the decorated Ising models consisting of mixed spins have attracted 
a great deal of attention on account of much richer critical behaviour in comparison 
with their single-spin counterparts. Exact solutions of the mixed-spin Ising models 
on several decorated planar lattices have furnished a deeper insight into diverse 
attractive issues of statistical mechanics such as multiply reentrant phase transitions \cite{yama69,hori83,gonc84,sant86,jasc98,lack98,mata07}, multicompensation phenomenon \cite{jasc98,lack98,mata07,dakh98}, annealed disorder \cite{falk80,sant87,gonc91,cout93,corr94,cost95}, as well as, the effect of non-zero external magnetic field \cite{mash73,gonc86,cano06}. 
In addition, the mixed-spin Ising models on some decorated planar lattices can also 
be viewed as useful model systems for some ferromagnetic, ferrimagnetic, and 
metamagnetic molecular-based magnetic materials (see Refs. \cite{ohba00,okaw02} 
for excellent recent reviews).

Among the most convenient properties of the generalized decoration-iteration transformation 
one could mention its general validity, which means that this mapping transformation holds independently of the lattice spatial dimension to be considered. Unfortunately, the application
of decoration-iteration mapping was until lately basically restricted to one- and 
two-dimensional decorated lattices due to the lack of the exact solution of the spin-1/2 
Ising model on three-dimensional (3D) lattices. The majority of studies 
concerned with the mixed-spin Ising models on 3D decorated lattices were therefore 
based on approximative analytical methods such as mean-field and effective-field theories \cite{kane97,kane98,sous00,kane01,aoua01,mout02,tava07}. On the other hand, essentially 
exact results were recently reported by Oitmaa and Zheng \cite{oitm03} for phase diagrams 
of the mixed-spin Ising model on the decorated cubic lattice by adopting the decoration-iteration transformation and the critical temperature of the corresponding spin-1/2 Ising model 
on the simple cubic lattice, which is known with a high numerical precision from 
the high-temperature series expansion \cite{bute97}. Another possibility of how rather 
accurate results can be obtained for the mixed-spin Ising model on 3D decorated lattices 
is to perform extensive Monte Carlo simulation as recently done by Boughrara and Kerouad for 
the decorated Ising film \cite{boug08}.

In the present work, the mixed spin-1/2 and spin-$S$ Ising model on the layered 3D decorated 
lattice will be studied by applying the decoration-iteration transformation, which establishes 
a precise mapping relationship with the spin-1/2 Ising model on the tetragonal lattice.
The reasonable results for the mixed-spin Ising model on the 3D decorated lattice can be 
consequently extracted from the corresponding results of much simpler spin-1/2 Ising model 
on the tetragonal lattice. Two alternative approaches are subsequently used for a theoretical 
analysis of the latter model: the first analytical approach is based on the Zhang's conjectured solution for the spin-1/2 Ising model on the orthorhombic lattice \cite{zhan07}, while the second 
numerical approach exploits Monte Carlo simulations. Even though there are serious doubts \cite{wu08a,wu08b,perk09} about a rigour of the conjectured solution for the spin-1/2 Ising model 
on the 3D orthorhombic lattice \cite{zhan07,zhan08,zhan09,klei08}, it is quite tempting to utilize 
it for a theoretical treatment of highly anisotropic spin systems because the Zhang's results \cite{zhan07} correctly reproduce the Onsager's exact solution for the spin-1/2 Ising model on the 2D rectangular lattice \cite{onsa44}. From this point of view, one should expect only small numerical error when treating highly anisotropic quasi-1D or quasi-2D spin systems even if the conjectured solution does not represent the true exact solution and moreover, the correctness of obtained results 
can easily be checked by the alternative numerical method based on the Monte Carlo simulations. 
The main advantage of the combination of the generalized decoration-iteration transformation 
with the Zhang's conjectured solution is that it preserves the analytic form of the solution 
to be obtained for the layered Ising model of mixed spins. This advantage is naturally lost 
in the case of combining the decoration-iteration transformation with Monte Carlo simulations.

The outline of this paper is as follows. In Section \ref{model}, the detailed description 
of the layered mixed-spin Ising model is presented at first. Then, some details of 
the decoration-iteration mapping are clarified together with two alternative ways of how 
the magnetization and critical temperature can be calculated. The most interesting results 
are presented and detailed discussed in Section \ref{result}. Finally, some concluding remarks 
are mentioned in Section \ref{conc}. 

\section{Ising model and its solution}
\label{model}

\begin{figure}
\begin{center}
\includegraphics[width=0.85\textwidth]{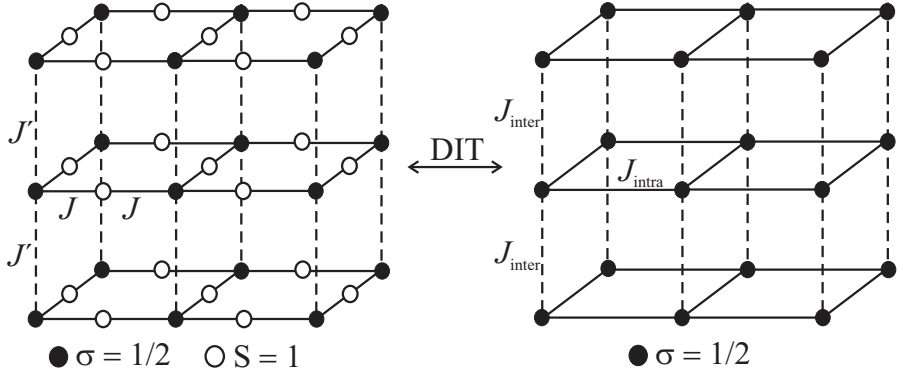}
\end{center}
\vspace{-0.2cm}
\caption{Schematic representation of the mixed spin-1/2 and spin-$S$ Ising model on the layered 3D decorated lattice and its decoration-iteration transformation towards the simple spin-1/2 Ising model on the tetragonal lattice. Solid (empty) circles denote lattice positions of the spin-1/2 (spin-$S$) atoms, while solid and broken lines represent intra- and inter-layer interactions 
for both mixed-spin as well as effective spin-1/2 Ising model, respectively.}
\label{fig1}
\end{figure}

Let us define the mixed spin-1/2 and spin-$S$ ($S \geq 1$) Ising model on the 3D layered decorated lattice as it is diagrammatically depicted in Fig.~\ref{fig1}. In this figure, the solid circles 
denote lattice positions of the spin-1/2 Ising atoms that reside sites of the simple cubic lattice 
and the empty ones represent lattice positions of the decorating spin-$S$ Ising atoms lying on the horizontal bonds of the simple cubic lattice. 
Let us further denote the total number of layers by the symbol $L$, the total number of the spin-1/2 atoms within each layer as $N = L \times L$ and the total number of the spin-1/2 atoms as $N_T = L \times L \times L$. The model under investigation can be then defined through the Hamiltonian
\begin{eqnarray}
H = - J \sum_{l=1}^{L} \sum_{(i,j)}^{4N} S_{l,i} \sigma_{l,j} 
    - J' \sum_{l=1}^{L} \sum_{j=1}^{N} \sigma_{l,j} \sigma_{l+1,j}
    - D \sum_{l=1}^{L} \sum_{i=1}^{2N} S_{l,i}^2,
\label{eq1}
\end{eqnarray}
where $\sigma_{l,j} = \pm 1/2$ and $S_{l,i} = -S, -S+1, \ldots, S$ are two different kinds of 
Ising spins located in the $l$th layer at $j$th and $i$th lattice position, respectively,
and periodic boundary conditions are imposed for simplicity. 
The parameter $J$ denotes the intra-layer interaction between the nearest-neighbour spin-1/2 
and spin-$S$ atoms, the parameter $J'$ labels the inter-layer interaction between the nearest-neighbour spin-1/2 atoms from two adjacent layers and the parameter $D$ stands for axial zero-field splitting (AZFS) parameter that acts on the decorating spin-$S$ atoms only \cite{rudo08a,rudo08b}. 

The partition function of the layered mixed-spin Ising model, which is defined through the Hamiltonian (\ref{eq1}), can be written after straightforward re-arrangement of some terms in the form
\begin{eqnarray}
Z &=& \sum_{\{ \sigma_{l,j} \}} \exp \left( \beta J' \sum_{l=1}^{L} 
          \sum_{j=1}^{N} \sigma_{l,j} \sigma_{l+1,j} \right) \nonumber \\
  &&  \times \prod_{l=1}^{L} \prod_{i=1}^{2N} \sum_{S_{l,i}=-S}^{S} 
  \exp \left[ \beta J S_{l,i} \left(\sigma_{l,i1} + \sigma_{l,i2} \right) + \beta D S_{l,i}^2 \right],  
\label{eq2}
\end{eqnarray}
where $\beta = 1/(k_{\rm B} T)$, $k_{\rm B}$ is Boltzmann's constant, $T$ is the absolute
temperature and the symbol $\sum_{\{ \sigma_{l,j} \}}$ stands for a summation over all possible
spin configurations of the spin-1/2 atoms. It can be readily seen from the structure of 
the relation (\ref{eq2}) that the summation over spin degrees of freedom of the decorating 
spin-$S$ atoms can be performed independently of each other (there is no direct interaction 
between the decorating spins) and before summing over all possible spin configurations of 
the spin-1/2 atoms. Both these facts enable us to introduce the generalized decoration-iteration transformation \cite{syoz72,fish59} 
\begin{eqnarray}
\sum_{S_{l,i}=-S}^{S} \exp [\beta J S_{l,i} (\sigma_{l,i1} + \sigma_{l,i2}) + \beta D S_{l,i}^2]
 = A \exp (\beta J_{\rm intra} \sigma_{l,i1} \sigma_{l,i2}),   
\label{eq3}
\end{eqnarray}
which effectively replaces all the interaction terms associated with the decorating spin 
$S_{l,i}$ and substitutes them by the equivalent expression that depends solely on its two nearest-neighbour vertex spins $\sigma_{l,i1}$ and $\sigma_{l,i2}$. Of course, the decoration-iteration transformation must retain its validity regardless of possible spin states of both the nearest-neighbour vertex spins $\sigma_{l,i1}$ and $\sigma_{l,i2}$ and this "self-consistency" condition unambiguously determines until now not specified transformation parameters $A$ and 
$J_{\rm intra}$
\begin{eqnarray}
A &=& \left \{  \left[ \sum_{n=-S}^S \exp(\beta D n^2) \cosh(\beta J n) \right] 
                \left[ \sum_{n=-S}^S \exp(\beta D n^2) \right] \right \}^{1/2}, 
\label{eq4a} \\ 
\beta J_{\rm intra} &=& 2 \ln \left[ \sum_{n=-S}^S \exp(\beta D n^2) \cosh(\beta J n) \right] 
                     -  2 \ln \left[ \sum_{n=-S}^S \exp(\beta D n^2) \right].
\label{eq4b}
\end{eqnarray}
At this stage, the substitution of the decoration-iteration transformation (\ref{eq3}) into Eq.~(\ref{eq2}) yields, after straightforward re-arrangement of few terms, the following mapping relationship for the partition function
\begin{eqnarray}
Z (\beta, J, J', D) = A^{2 N L} Z_{\rm tetragonal} (\beta, J_{\rm intra}, J_{\rm inter}=J').  
\label{eq5}
\end{eqnarray}
It is quite obvious that the mapping relation (\ref{eq5}) relates the partition function 
of the layered Ising model on 3D decorated lattice to that one of the corresponding 
spin-1/2 Ising model on the tetragonal lattice (see Fig.~\ref{fig1}). Notice furthermore 
that the effective intra-layer interaction $J_{\rm intra}$ of the corresponding spin-1/2 
Ising model on the tetragonal lattice is temperature dependent parameter satisfying 
the self-consistency condition (\ref{eq4b}), while the effective inter-layer interaction 
$J_{\rm inter}$ is temperature independent parameter that is directly equal to 
the interaction parameter $J'$. 

A calculation of the spontaneous magnetization and other thermodynamic quantities 
can be now accomplished in an easy and rather straightforward way. Adopting the mapping 
theorems developed by Barry \textit{et al}. \cite{barr88,barr90,barr91,barr95}, 
the sublattice magnetization $m_{\rm A}$ relevant to the spin-1/2 atoms of the mixed-spin 
Ising model on 3D decorated lattice directly equals to the magnetization of 
the corresponding spin-1/2 Ising model on the tetragonal lattice
\begin{eqnarray}
m_{\rm A} (\beta, J, J', D) \equiv \langle \sigma_{l,i} \rangle_{\rm decorated} 
                        = \langle \sigma_{l,i} \rangle_{\rm tetragonal} 
                          \equiv m_{0} (\beta, J_{\rm intra}, J_{\rm inter}).                          
\label{eq6}
\end{eqnarray}
Above, the symbols $\langle \ldots \rangle_{\rm decorated}$ and $\langle \ldots \rangle_{\rm tetragonal}$ denote canonical ensemble averaging performed within the mixed-spin Ising model 
on the 3D decorated lattice and its corresponding spin-1/2 Ising model on the tetragonal lattice, respectively. On the other hand, the sublattice magnetization $m_{\rm B}$ of the spin-$S$ atoms 
can easily be calculated by combining the exact Callen-Suzuki spin identity \cite{call63,suzu65} 
with the differential operator technique \cite{honm79,kane93}. It is noteworthy that this kind 
of mathematical treatment essentially follows Kaneyoshi's procedure \cite{kane96} originally 
developed for the decorated planar Ising models, which connects the sublattice magnetization 
of the spin-$S$ atoms with that one of the spin-1/2 atoms through the relation
\begin{eqnarray}
m_{\rm B} \equiv  \langle S_{l,i} \rangle_{\rm decorated} 
        = 2 m_{\rm A} \frac{\displaystyle \sum_{n=-S}^S n \exp(\beta D n^2) \sinh(\beta J n)}
                           {\displaystyle \sum_{n=-S}^S   \exp(\beta D n^2) \cosh(\beta J n)}.   
\label{eq7}
\end{eqnarray}
If both sublattice magnetization are known, the total magnetization of the mixed-spin Ising model 
on the 3D decorated lattice is given by the definition $m=(m_{\rm A} + 2m_{\rm B})/3$.

It is quite obvious from Eqs.~(\ref{eq6}) and (\ref{eq7}) that it is now sufficient to find 
the spontaneous magnetization of the corresponding spin-1/2 Ising model on the tetragonal lattice 
in order to complete our calculation of both sublattice magnetizations. For this purpose, we will utilize two alternative approaches: the first method adopts the conjectured solution for the spin-1/2 Ising model on the orthorhombic lattice \cite{zhan07}, while the second method takes advantage of  
numerical Monte Carlo simulations. The former analytic procedure employs an explicit expression 
for the spontaneous magnetization of the spin-1/2 Ising model on the tetragonal lattice, which
can be easily descended from the Zhang's results for the spin-1/2 Ising model on the orthorhombic lattice \cite{zhan07}
\begin{eqnarray}
m_{0} = \frac{1}{2} \left[ \frac{(1 - x^2 - x^2 y^4 + x^4 y^4)^2 - 16 x^4 y^4}
                                {(1 - x^2)^2 (1 - x^2 y^4)^2} \right]^{3/8},  
\label{eq8}
\end{eqnarray}
where $x = \exp(- \beta J_{\rm intra}/2)$ and $y = \exp(- \beta J_{\rm inter}/2)$. Within the framework of this analytic method, it is also easy to obtain the critical condition that thoroughly determines 
a critical point of the order-disorder phase transition of the layered Ising model on the 3D decorated lattice. Namely, both sublattice magnetization $m_{\rm A}$ and $m_{\rm B}$ tend necessarily 
to zero if the spontaneous magnetization $m_0$ of the corresponding spin-1/2 Ising model 
on the tetragonal lattice vanishes as well. Accordingly, the critical condition that enables 
to locate the order-disorder phase transition of the mixed-spin Ising model on 3D decorated lattice
can readily be found from the Zhang's critical condition for the spin-1/2 Ising model on the orthorhombic lattice \cite{zhan07}, which contains as a particular case the following critical condition 
for the spin-1/2 Ising model on the tetragonal lattice 
\begin{eqnarray}
\sinh \left( \frac{\beta_c J_{\rm intra}}{2} \right) 
\sinh \left( \frac{\beta_c J_{\rm intra}}{2} +  \beta_c J_{\rm inter} \right) = 1,
\label{eq9}
\end{eqnarray}
where $\beta_c = 1/(k_{\rm B} T_c)$ and $T_c$ denotes the critical temperature. It should be nevertheless mentioned that the above critical condition thoroughly determines a critical behaviour 
of the layered Ising model of mixed spins on assumption that the effective intra-layer interaction $J_{\rm intra}$ satisfies the mapping relation (\ref{eq4b}) and the effective inter-layer interaction 
is equal to $J_{\rm inter} = J'$.

To avoid a danger of over-interpretation of the obtained results, the Monte Carlo simulations \cite{bind88,newm99} were further used as the other alternative approach with the aim to provide an independent calculation of the spontaneous magnetization of the corresponding spin-1/2 Ising model on the tetragonal lattice defined via the effective interactions $J_{\rm intra}$ and $J_{\rm inter}$.
The main advantage of a combination of the decoration-iteration transformation with the Monte Carlo method consists in a drastic reduction of the total Hilbert space, because the total number of available spin configurations reduces from $[2(2S+1)^2]^{N_T}$ to $2^{N_T}$ after performing 
the decoration-iteration transformation. Apparently, this drastic reduction of the total Hilbert 
space makes from the Monte Carlo simulations much more efficient tool for obtaining meaningful 
results. To be more specific, we have performed 
the Monte Carlo simulations for the spin-1/2 Ising model on the tetragonal lattice with the 
linear size $L = 10$, $20$, $30$, and $40$. Note furthermore that periodic boundary conditions 
were imposed and all initial spin states were randomly assigned. The last spin configuration 
at any temperature was used as an input for maintained simulation at lower temperature. 
Spin configurations were generated by random passing through the tetragonal lattice and 
making single spin-flip attemps, which were accepted or rejected according to 
the standard Metropolis algorithm \cite{metr53}. Finally, canonical ensemble averages were 
calculated using $10^6$ Monte Carlo steps per site after discarding the initial 
$2 \times 10^5$ Monte Carlo steps per site.

The magnetization per site was calculated from the definition $m_0 = \langle | m_{MC} | \rangle 
\equiv (1/L^3) \langle | \sum_{l,i} \sigma_{l,i} | \rangle_{L}$, where the symbol $\langle \dots \rangle_{L}$ denotes the ensemble average performed within the spin-1/2 Ising model on the 
tetragonal lattice with the linear size $L$. It is worthy to remind that the magnetization $m_0$ 
then directly equals to the sublattice magnetization $m_{\rm A}$ relevant to the spin-1/2 atoms 
of the mixed-spin Ising model on 3D decorated lattice. The other sublattice magnetization 
$m_{\rm B}$ of the spin-$S$ atoms can easily be enumerated from the relation (\ref{eq7}). 
For better accuracy, the critical temperature was determined with the help of fourth-order 
Binder cumulants $U_L = 1 - \langle m_{MC}^4 \rangle_L / [3 \langle m_{MC}^2 \rangle^2_L]$ \cite{bind81a,bind81b}, which intersect each other for different lattice sizes $L$ 
at a critical point according to the finite-size scaling theory \cite{bind88}.

\section{Results and discussion}
\label{result}

In this part, let us proceed to a discussion of the most interesting results obtained 
for the layered Ising model on 3D decorated lattice. Before doing this, it is worthy 
to mention that all analytical results presented in the preceding section are rather 
general as they hold for arbitrary quantum spin number $S$ of the decorating spins 
and also independently of whether ferromagnetic or antiferromagnetic interactions 
$J$ and $J'$ are assumed. In what follows, we will restrict ourselves for simplicity 
just to an analysis of the particular case with both ferromagnetic interaction constants 
$J>0$ and $J'>0$. It should be mentioned, however, that the presented zero-field phase 
diagrams should remain valid also for layered Ising models with the antiferromagnetic 
interaction(s) $J$ and/or $J'$ due to an invariance of Ising spin systems with respect 
to the transformations $J \to -J$ and/or $J' \to -J'$, which merely cause a rather 
trivial change of the ferromagnetic ($J>0,J'>0$) alignment to the metamagnetic ($J>0,J'<0$), 
the ferrimagnetic ($J<0,J'>0$), or the antiferromagnetic ($J<0,J'<0$) one. 

\begin{figure}
\begin{center}
\includegraphics[width=0.95\textwidth]{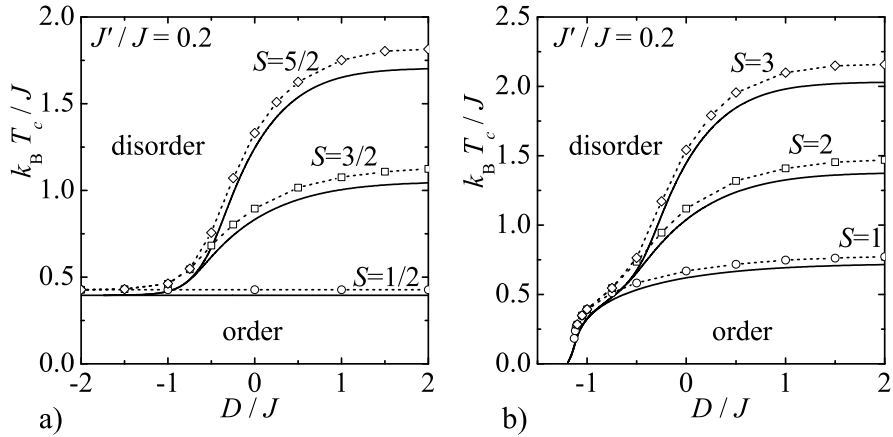}
\end{center}
\vspace{-1cm}
\caption{The critical temperature as a function of the AZFS parameter $D/J$ for several values 
of the decorating spins $S$ when the ratio between the inter- and intra-layer interactions 
is fixed to $J'/J = 0.2$. Solid lines depict the critical temperatures calculated from 
Eq.~(\ref{eq9}) of our analytical procedure, whereas symbols connected by dotted lines 
show the corresponding numerical results obtained by using Monte Carlo simulations.}
\label{fig2}
\end{figure}

First, let us take a closer look at finite-temperature phase diagrams, which are shown in Fig.~\ref{fig2} in the form of the critical temperature vs. the AZFS parameter dependences 
for several values of the decorating spins $S$ and the selected ratio $J'/J = 0.2$.
In this figure, the solid lines depict analytical results obtained from the critical 
condition (\ref{eq9}), while symbols connected by dotted lines show the corresponding numerical 
results acquired by the use of Monte Carlo simulations. It is quite obvious from Fig.~\ref{fig2} 
that the phase diagrams obtained from both independent theoretical approaches are in a good 
qualitative agreement, the numerical data for critical temperatures stemming from Monte Carlo simulations are in fact just slightly above the respective analytical results. 
Moreover, it can be also clearly seen from Fig.~\ref{fig2} that the overall critical behaviour 
basically depends merely on whether the decorating spins are half-odd-integer or integer ones. 
The critical temperature for the spin systems with half-odd-integer decorating spins 
(Fig.~\ref{fig2}a) monotonically decreases upon decrease of the AZFS parameter until it 
asymptotically reaches the critical temperature of the special case with $S=1/2$ that is 
of course independent of the AZFS parameter. This rather trivial finding can be straightforwardly attributed to a consecutive lowering of the spin state of the half-odd-integer decorating spins, 
which generally takes place at sufficiently strong negative values of the AZFS parameters 
$D/J = -1/(2n)$ on assumption that the relevant spin state changes from $S_{l,i}=n+1/2$ 
to $S_{l,i}=n-1/2$ ($n = 1,2,3,\ldots$). 
Similarly, the critical temperature for the spin systems with integer decorating spins (Fig.~\ref{fig2}b) monotonically decreases upon decrease of the AZFS parameter until 
it tends towards zero temperature at some boundary value of the AZFS parameter. 
The monotonous decrease of the critical temperature can be again explained in terms of 
a gradual decline of the spin state of integer decorating spins, which takes place 
at the following values of the AZFS parameter $D/J = -1/(2n-1)$ provided that the spin state 
changes from $S_{l,i}=n$ to $S_{l,i}=n-1$ ($n = 1,2,3, \ldots$). Contrary to our expectations, 
the critical temperatures of the spin systems with integer decorating spins do not vanish 
at the boundary value of the AZFS parameter, $D/J = -1$, below which all integer decorating 
spins tend towards their 'non-magnetic' spin state $S_{l,i}=0$. This is the most remarkable
finding of our study and we will henceforth explore this striking critical behaviour, which 
represents a general feature of the spin systems with integer decorating spins, 
on the simplest model with the integer decorating spins $S=1$. 

Let us consider first possible spin arrangements to emerge in the ground state 
of this particular model system. 
It turns out that three different phases may appear in total at zero temperature 
in dependence on a relative strength of the intra-layer interaction $J$, 
the inter-layer interaction $J'$, and the AZFS parameter $D$. The AZFS term $D$ plays 
the role of the anisotropy parameter that forces all decorating spins $S=1$ towards 
their 'non-magnetic' spin state $S_{l,i}=0$ provided that this parameter is a sufficiently 
large negative number. The usual ferromagnetic phase (FP), which can be characterized 
through the following spin states of the decorating and vertex spins 
$(S_{l,i}; \sigma_{l,i}) = (1; 1/2)$, consequently represents the lowest-energy state 
just if $D/J>-1$. Note that this finding is consistent with the relevant results of 
our analytical approach as well as the numerical Monte Carlo simulations. On the other hand, 
it directly follows from the critical condition (\ref{eq9}) that the striking 'quasi-1D' 
ferromagnetic phase (QFP) constitutes the ground state in a range of intermediate strong 
anisotropy parameters $D/J \in (-1-J'/J,-1)$, where it exhibits an outstanding spontaneous 
long-range order unambiguously determined through the spin states 
$(S_{l,i}; \sigma_{l,i}) = (0; 1/2)$. It should be stressed that the qualitatively same behaviour 
is also predicted by Monte Carlo simulations even although it becomes rather hard to estimate accurately the lower boundary of QFP within this numerical technique (see for details 
the subsequent part). The absence of any spontaneous long-range order can finally be detected 
in the disordered phase (DP), which becomes the lowest-energy state on assumption that $D/J<-1-J'/J$ 
when the critical condition (\ref{eq9}) is taken into account. In this particular case, 
the sufficiently strong (negative) AZFS parameter energetically favours the 'non-magnetic' 
spin state $S_{l,i}=0$ of the decorating spins and hence, there appears the spin state $(S_{l,i}; \sigma_{l,i}) = (0; \pm 1/2)$ with a complete randomness in the states of the vertex spins 
(the vertex spins from the same layer do not effectively feel each other). The most surprising 
finding resulting from our study of the ground state is a pure existence of QFP, which exhibits 
a remarkable spontaneous long-range order in spite of the 'non-magnetic' nature of 
the decorating spins and the effectively 'quasi-1D' character of the spin system.

\begin{figure}
\begin{center}
\includegraphics[width=0.95\textwidth]{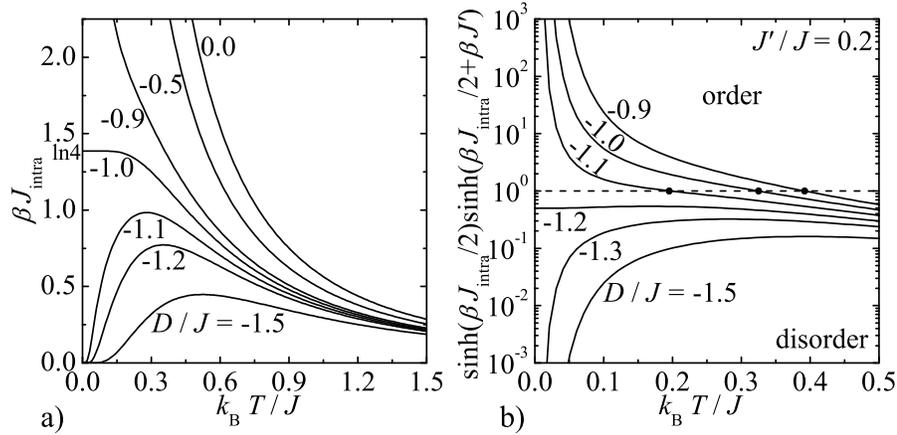}
\end{center}
\vspace{-1cm}
\caption{Typical temperature dependences of the effective intra-layer coupling $\beta J_{\rm intra}$ 
are shown in Fig.~\ref{fig3}a) for several values of the AZFS parameter $D/J$. 
Note that $\beta J_{\rm intra}$ is given by the mapping relation (\ref{eq4b}) and it does not 
depend on a strength of the inter-layer interaction $J'$. Fig.~\ref{fig3}b) displays 
in a semi-logarithmic scale a graphical solution of the critical condition (\ref{eq9}). 
Solid (broken) lines depict temperature dependences of the left-hand-side (right-hand-side) 
of the critical condition (\ref{eq9}) for several values of the AZFS parameter $D/J$ and 
the ratio $J'/J = 0.2$. The points of intersection between broken and solid lines 
(full circles) determine critical points.}
\label{fig3}
\end{figure}

To provide a deeper insight into the mechanism that drives the spin system into one of 
those three available spin states, it might be useful to take a closer look at the effective 
coupling parameters $\beta J_{\rm intra}$ and $\beta J_{\rm inter}$ of the corresponding spin-1/2 
Ising model on the tetragonal lattice, which were used both in our analytical approach 
as well as Monte Carlo simulations. The effective inter-layer coupling $\beta J_{\rm inter} 
= J'/(k_{\rm B} T)$ is evidently monotonously decreasing function of the temperature, 
which diverges as $T^{-1}$ when approaching the zero temperature. By contrast, 
the effective intra-layer coupling $\beta J_{\rm intra}$ exhibits much more complex 
thermal variations, which are for better illustration depicted in Fig.~\ref{fig3}a) 
for several values of the AZFS parameter $D/J$. It can be directly proved from 
the definition (\ref{eq4b}) that $\beta J_{\rm intra}$ diverges as $T^{-1}$ when
reaching the zero temperature either according to the law $\beta J_{\rm intra} = 2J/(k_{\rm B} T)$ valid for $D/J>0$, or according to the formula $\beta J_{\rm intra} = 2(D+J)/(k_{\rm B} T)$ 
valid for $D/J \in (-1,0)$. Furthermore, the effective intra-layer coupling tends 
towards the constant value $\beta J_{\rm intra} = \ln 4$ when approaching zero temperature 
for the special case $D/J=-1$, while it exponentially goes to zero by following the law 
$\beta J_{\rm intra} = 2 \exp[(D+J)/(k_{\rm B} T)]$ in the region $D/J<-1$. Notice that 
all aforedescribed features can also be clearly seen in the dependences shown in Fig.~\ref{fig3}a).
This comprehensive analysis of the effective intra-layer coupling demonstrates that 
there does not exist (at least at zero temperature) any effective intra-layer interaction 
between the spin-1/2 atoms if $D/J<-1$ and thus, the spin-1/2 atoms from the same layer should 
become completely independent of each other under this condition. This reasoning would have 
a simple physical explanation, since the relative strength of AZFS parameter $D/J=-1$ is 
just as strong as to make energy balance between the 'non-magnetic' ($S_{l,i}=0$) and 
magnetic ($S_{l,i}=1$) spin state of the decorating spins and accordingly, all vertex spins
should be effectively separated by the 'non-magnetic' decorating spins $S_{l,i}=0$ whenever $D/J<-1$. 

Bearing all this in mind, one would intuitively expect that the layered Ising model on 3D decorated 
lattice must be disordered at any finite temperature when $D/J<-1$. Under this assumption, the only non-zero term at the zero temperature is the effective inter-layer interaction $J_{\rm inter}=J'$ 
and the layered Ising model on 3D decorated lattice should therefore break into a set of the independent spin-1/2 Ising chains (running perpendicular to the layers) that do not possess 
a finite critical temperature. However, the mathematical structure of the critical condition (\ref{eq9}) as well as the numerical results from Monte Carlo simulations indicate a more involved situation. Our analytical approach implies that the spin system is spontaneously ordered 
(disordered) if the product on the left-hand-side of the critical condition (\ref{eq9}) 
is greater (less) than unity. 
Thus, there exists a possibility that the product on the left-hand-side of the critical 
condition (\ref{eq9}) might be greater than unity despite the zero value of the effective intra-layer coupling, for instance, if a divergence of the effective inter-layer coupling $\beta J_{\rm inter}$ 
overwhelms the asymptotic vanishing of the intra-layer coupling  $\beta J_{\rm intra}$. 
One actually finds in the zero temperature limit ($T \to 0$ or equivalently $\beta \to \infty$) that 
\begin{eqnarray*}
\lim_{\beta \to \infty} \left [ 
\sinh \left(\frac{\beta J_{\rm intra}}{2} \right) 
\sinh \left( \frac{\beta J_{\rm intra}}{2} + \beta J_{\rm inter} \right) \right ]
=  \biggl \{ \begin{array}{ll} 
\infty  & \, {\rm if} \, \, \, \frac{D}{J} > - 1 - \frac{J'}{J}  \\
 \: 0   & \, {\rm if} \, \, \, \frac{D}{J} < - 1 - \frac{J'}{J} 
         \end{array},  
\end{eqnarray*}
which means that the spontaneous order disappears only at $D/J = - 1 - J'/J$ notwithstanding 
the simple intuitive expectations given above. Among other matters, this argument might serve 
in evidence of the outstanding spontaneous long-range ordering QFP that emerges in a range 
of the intermediate strong anisotropy parameters $D/J \in (-1-J'/J,-1)$ despite the 
'non-magnetic' nature of all decorating spins. For better illustration, Fig.~\ref{fig3}b) 
shows in a graphical form several temperature dependences of the left-hand-side of 
the critical condition (\ref{eq9}) for one particular value of the ratio $J'/J=0.2$, 
which confirm a correctness of the aforedescribed analysis. It is noteworthy that this figure 
can also be regarded as a graphical solution of the critical condition (\ref{eq9}) that determines 
a critical point of the layered Ising model on 3D decorated lattice as an intersection
of both sides of the Eq.~(\ref{eq9}). Finally, it is worth noticing that the above mentioned 
analysis is also consistent with the numerical results of Monte Carlo simulations, which 
predict the spontaneous order for intermediate values of the AZFS parameter $D/J \lesssim -1$ 
as well. 

\begin{figure}
\begin{center}
\includegraphics[width=0.6\textwidth]{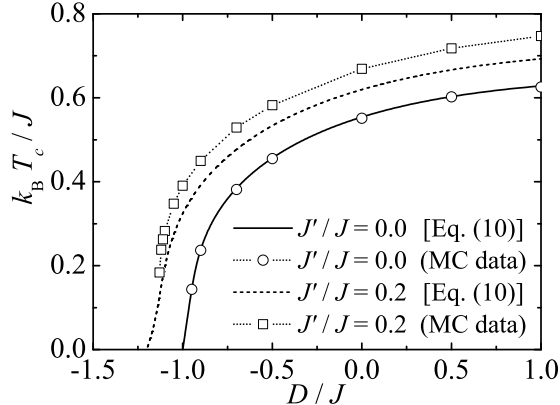}
\end{center}
\vspace{-0.9cm}
\caption{The critical temperature as a function of the AZFS parameter $D/J$ for the particular spin case $S=1$ and two different values of the ratio $J'/J = 0.0$ and $0.2$ between the inter- and intra-layer interactions. The solid and dashed lines without any symbol show the critical lines 
obtained from Eq.~(\ref{eq9}). The symbols connected by dotted lines display the relevant 
critical points acquired from Monte Carlo simulations.}
\label{fig4}
\end{figure}

For comparison, we depict in Fig.~\ref{fig4} the critical temperature as a function of the AZFS parameter for the particular spin case $S=1$ and two different values of the interaction ratio 
$J'/J = 0.0$ and $0.2$. The critical temperatures, which are displayed in Fig.~\ref{fig4} as 
solid and dashed lines, were obtained by numerically solving the critical condition (\ref{eq9}). 
The symbols connected by dotted lines depict the relevant numerical data obtained by using 
Monte Carlo simulations. 
It is worthwhile to remark that the critical line displayed in Fig.~\ref{fig4} for the special case $J'/J = 0$ is fully consistent with the formerly published exact results \cite{jasc98,dakh98}. 
In this particular case, the critical line actually ends up at the expected ground-state boundary $D/J=-1$ at which the spin state $S_{l,i}=1$ changes to the 'non-magnetic' one $S_{l,i}=0$ and 
there does not appear a striking spontaneous order inherent to QFP. This is a direct consequence 
of the fact that the critical condition (\ref{eq9}) extracted from the Zhang's solution for the spin-1/2 Ising model on the orthorhombic lattice \cite{zhan07} essentially reduces to the famous Onsager's solution for the spin-1/2 Ising model on the square lattice \cite{onsa44}. It should be pointed out, moreover, 
that the relevant numerical data from Monte Carlo simulations are lying on this critical line, 
which confirms accuracy of our Monte Carlo simulations. The critical temperature monotonically decreases with a decrease of the AZFS parameter also for any non-zero inter-layer interaction 
$J'/J \neq 0$ until it tends to zero at some stronger (more negative) values of the AZFS parameter 
(see the curve for the particular case $J'/J = 0.2$). This surprising 
finding is evident both from our analytical results as well as Monte Carlo simulations. However, 
the decorating spins reside the spin state $S_{l,i}=1$ just if $D/J > - 1$, while they reside the 'non-magnetic' spin state $S_{l,i}=0$ whenever $D/J < - 1$. From this perspective, the boundary 
value of the AZFS parameter $D/J = -1$ divides the critical line into two different region: 
the part where $D/J > - 1$ corresponds to the critical points of the FP, while the part where 
$D/J < - 1$ corresponds to the critical points of the QFP. It should be also mentioned that 
the lower boundary allocating a presence of the spontaneously ordered QFP is $D/J = -1 - J'/J$ according to the critical condition (\ref{eq9}), while it becomes rather hard to locate 
precisely the lower boundary with the help of Monte Carlo simulations. Namely, the more and more extensive Monte Carlo simulations are needed at sufficiently low temperatures in order 
to overcome finite-size effects that become very important in the parameter space $D/J < -1$, 
because the relevant spin system effectivelly splits into a set of weakly interacting 
spin-1/2 Ising chains running perpendicular to the layers. 

\begin{figure}
\begin{center}
\includegraphics[width=0.95\textwidth]{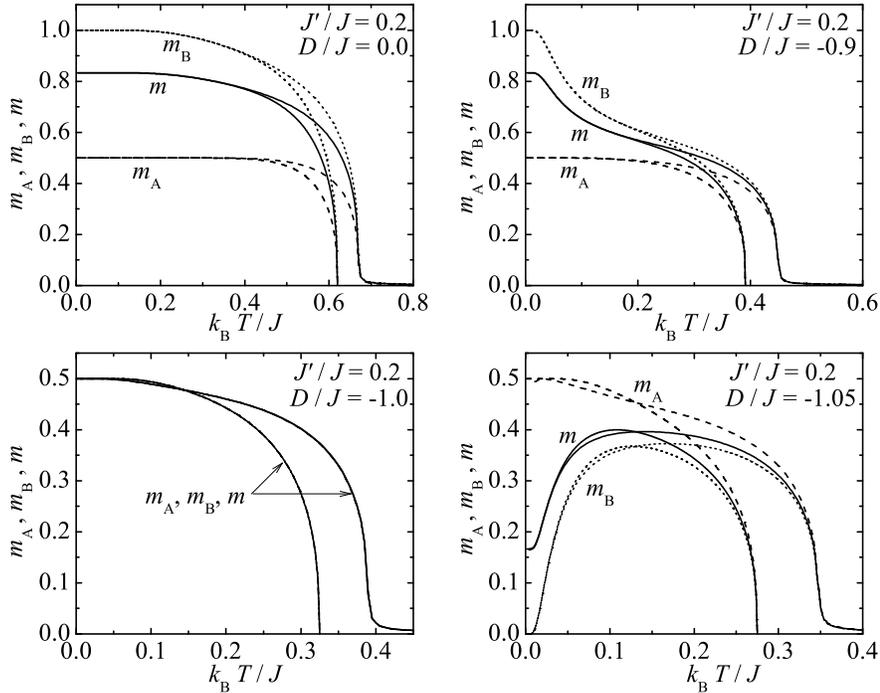}
\end{center}
\vspace{-1.1cm}
\caption{Thermal dependences of the total magnetization $m$ (solid lines) and the sublattice magnetization $m_{\rm A}$ (dashed lines), $m_{\rm B}$ (dotted lines) for the spin case $S=1$, 
the fixed value of the ratio $J'/J = 0.2$ and several values of the AZFS parameter $D/J$. 
The lines ending at lower critical temperature show the relevant results of our analytical 
approach, while the lines having a high-temperature tail come from the Monte Carlo simulations 
of a lattice with the linear size $L=40$.}
\label{fig5}
\end{figure}

To provide an independent check of a presence of spontaneously ordered QFP, it might be quite 
useful to take a look at thermal dependences of the total and sublattice spontaneous magnetizations. 
For this purpose, some temperature variations of the total and sublattice magnetizations are displayed 
in Fig.~\ref{fig5} for the particular value of the interaction ratio $J'/J=0.2$ and several values
of the anisotropy parameter $D/J$. It is noteworthy that the results obtained from our analytical procedure are in a good qualitative accordance with the numerical estimates of Monte Carlo simulations. However, there appears just a small deviation between the relevant results at relatively
high temperatures close to a critical point, because our analytical procedure slightly underestimates
the critical temperature in comparison with the Monte Carlo predictions. 
Fig.~\ref{fig5}a) shows thermal dependences of the total and sublattice magnetizations, which are typical for $D/J \gtrsim 0$ and which lead to the most common Q-type temperature dependence of 
the total magnetization.
On the other hand, the S-type temperature dependence of the total magnetization can be observed 
on assumption that the AZFS parameter is slightly greater than the boundary value $D/J=-1$ [see 
Fig.~\ref{fig5}b)  for $D/J=-0.9$]. The stair-like S-shaped dependence with a rapid 
initial decrease of the total magnetization obviously appears owing to preferred thermal 
excitations of the decorating spins to the 'non-magnetic' spin state $S_{l,i}=0$. Namely,  
these thermal excitations are also reflected in the temperature dependence of the sublattice magnetization $m_{\rm B}$ and the 'non-magnetic' spin state $S_{l,i}=0$ is close enough 
in energy to the spin state $S_{l,i}=1$ to emerge in the ground state under this condition. Interestingly, 
the standard thermal dependences of Q-type are recovered for the total and both sublattice magnetizations by selecting the boundary value $D/J=-1$ (see Fig.~\ref{fig5}c). It is 
worthwhile to remark, nevertheless, that the sublattice magnetization $m_{\rm B}$ pertinent 
to the decorating spins starts in this particular case from one half of its saturation value 
on behalf of the energetic equivalence between the spin states $S_{l,i}=0$ and $S_{l,i}=1$, 
which are populated with the same probability. As a result, both sublattice magnetization 
exhibit the qualitatively same dependences that cannot be distinguished within the displayed 
scale. Last but not least, the interesting L-type 
dependence of the total magnetization can be found for the AZFS parameters $D/J<-1$ 
as depicted in Fig.~\ref{fig5}d) for the particular case $D/J=-1.05$. As one can see from 
this figure, the sublattice magnetization $m_{\rm B}$ of the decorating spins starts from 
zero and this might be regarded as another convincing evidence of the existence QFP. Besides, 
the temperature-induced increase of the total magnetization evidently comes from the relevant 
thermal excitations of the decorating spins, which are clearly reflected in the thermal behaviour 
of the sublattice magnetization $m_{\rm B}$. In agreement with this suggestion, the observed temperature-induced increase of the magnetization is the more robust, the closer is 
the AZFS parameter to the boundary value $D/J=-1$, i.e. the closer in energy is 
the excited magnetic spin state $S_{l,i}=1$ to the 'non-magnetic' spin state 
$S_{l,i}=0$ emerging at $T=0$.

\section{Conclusions}
\label{conc}

In the present work, the critical behaviour and magnetic properties of the layered Ising model 
of mixed spins on 3D decorated lattice are investigated by the use of generalized decoration-iteration transformation, which establishes a precise mapping relationship between the investigated model system 
and the corresponding spin-1/2 Ising model on the tetragonal lattice. This exact mapping method 
was subsequently combined either with the conjectured solution for the spin-1/2 Ising model on 
the orthorhombic lattice \cite{zhan07} or numerical Monte Carlo simulations with the aim to obtain 
the meaningful results for the mixed-spin Ising model on the layered 3D decorated lattice. 
The main advantage of the former procedure is that it preserves essentially analytical form
of the results obtained for critical and thermodynamic properties of the layered mixed-spin 
Ising model, while the main advantage of the latter procedure rest in a drastic reduction 
of the total Hilbert space that makes Monte Carlo simulations very efficient. In the spirit 
of both these techniques, the ground-state and finite temperature phase diagrams have been studied along with possible temperature dependences of the total and sublattice magnetizations.

The most interesting finding presented in this work surely represents a theoretical prediction 
of the striking spontaneous long-range ordering QFP, which appears in spite of the 'non-magnetic' nature of all decorating spins and the effectively 'quasi-1D' character of the spin system.
It should be pointed out, however, that the analogous spontaneous long-range order of the 
effectively 'quasi-1D' spin system have already been exactly confirmed in the mixed-spin Ising 
model on a decorated square lattice with two different kinds of decorating spins on the horizontal 
and vertical bonds \cite{stre07,cano08}. This noticeable and rather surprising coincidence 
can readily be understood from the mathematical structure of the critical condition (\ref{eq9}).
Indeed, the proposed critical condition (\ref{eq9}) for the spin-1/2 Ising model on the tetragonal lattice formally coincides with the Onsager's critical condition \cite{onsa44} derived for 
the spin-1/2 Ising model on the anisotropic square (rectangular) lattice to which the mixed-spin 
Ising model on anisotropically decorated square lattice is effectively mapped \cite{stre07,cano08}.

Finally, it is worthwhile to remark that the presented solution can be rather straightforwardly 
extended to account for several additional interaction terms not included in the Hamiltonian (\ref{eq1}) such as the biaxial zero-field splitting parameter acting on the decorating spins, 
the next-nearest-neighbour interaction between the vertex spins, the multispin interaction 
between the decorating spin and its two nearest-neighbour vertex spins and so on. 
 
\begin{ack}
The authors would like to thank Dr. Z.-D.~Zhang and Dr. J.H.H.~Perk for several useful comments
on the conjectured solution for the spin-1/2 Ising model on the orthorhombic lattice \cite{zhan07}.
\end{ack}

\end{document}